\documentclass[prb,twocolumn,showpacs,floatfix]{revtex4}

\usepackage{graphicx}
\usepackage{bm}

\begin{document}

\title{Angular magnetoresistance oscillations in quasi-one-dimensional
organic conductors in the presence of a crystal superstructure}

\author{Anand Banerjee}
\affiliation{Joint Quantum Institute and Center for Nanophysics and
Advanced Materials, Department of Physics, University of Maryland,
College Park, Maryland 20742-4111, USA}

\author{Victor M. Yakovenko}
\affiliation{Joint Quantum Institute and Center for Nanophysics and
Advanced Materials, Department of Physics, University of Maryland,
College Park, Maryland 20742-4111, USA}

\date{\bf cond-mat/0608317, v.1 August 14, 2006, v.3 October 2, 2008}


\begin{abstract}
We study the effect of crystal superstructures produced by
orientational ordering of the $\rm ReO_4$ and $\rm ClO_4$ anions in
the quasi-one-dimensional organic conductors, $\rm(TMTSF)_2ReO_4$ and
$\rm(TMTSF)_2ClO_4$, on the angular magnetoresistance oscillations
(AMRO) observed in these materials.  Folding of the Brillouin zone due
to anion ordering generates effective tunneling amplitudes between
distant chains.  These amplitudes cause multiple peaks in interlayer
conductivity for the magnetic-field orientations along the rational
crystallographic directions (the Lebed magic angles).  Different wave
vectors of the anion ordering in $\rm(TMTSF)_2ReO_4$ and
$\rm(TMTSF)_2ClO_4$ result in the odd and even Lebed angles, as
observed experimentally.  When a strong magnetic field is applied
parallel to the layers and perpendicular to the chains and exceeds a
certain threshold, the interlayer tunneling between different branches
of the folded electron spectrum becomes possible, and interlayer
conductivity should increase sharply.  This effect can be utilized to
probe the anion ordering gaps in $\rm(TMTSF)_2ClO_4$ and
$\rm(TMTSF)_2ReO_4$.  An application of this effect to
$\kappa$-$\rm(ET)_2Cu(NCS)_2$ is also briefly discussed.
\end{abstract}


\maketitle

\section{Introduction}

The quasi-one-dimensional (Q1D) organic conductors $\rm(TMTSF)_2X$
(where TMTSF is tetramethyltetraselenafulvalene and X represents a
monovalent anion, such as $\rm PF_6$, $\rm ClO_4$, or $\rm ReO_4$)
have very interesting physical properties, including the quantum Hall
effect and possibly triplet superconductivity
\cite{book-Ishiguro,book-Lebed}.  These materials consist of parallel
conducting chains along the $x$ axis, arranged in layers with the
interchain spacing $b$ along the $y$ axis and the interlayer spacing
$c$ along the $z$ axis.  The electron-tunneling amplitudes between the
TMTSF molecular sites are highly anisotropic in the three directions:
$t_a:t_b:t_c=2500:250:10$ K \cite{book-Ishiguro}.

\begin{figure}[b]
\includegraphics[width=0.8\linewidth]{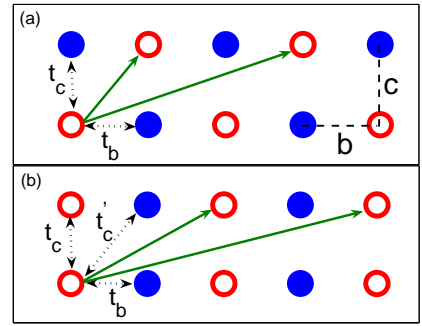}
\caption{(Color online) A view along the chains of a Q1D metal with
the anion ordering at a wave vector $\bm Q$. The filled and open
circles represent the chains with the energies $\pm E_g$. (a) $\rm
(TMTSF)_2ReO_4$ and $\bm Q=(0,1/2,1/2)$. (b) $\rm (TMTSF)_2ClO_4$ and 
$\bm Q=(0,1/2,0)$.}
\label{fig:lattice}
\end{figure}

These materials exhibit the angular magnetoresistance oscillations
(AMRO), where resistivity strongly changes as a function of the
magnetic-field orientation.  There are three basic types of AMRO: the
Lebed magic angles
\cite{Lebed86a,Osada91,Naughton91,Chaikin92c,Behnia94} for the
magnetic-field rotation in the $(y,z)$ plane, the Danner-Kang-Chaikin
(DKC) oscillations in the $(x,z)$ plane \cite{Chaikin94a,Chaikin95c},
and the third angular effect in the $(x,y)$ plane
\cite{Yoshino95,Osada96,Lebed97b}.  The Lebed oscillations manifest
themselves as sharp peaks in the interlayer conductivity $\sigma_{zz}$
occurring when the magnetic field points from one chain to another
along a rational crystallographic direction, as illustrated in Fig.\
\ref{fig:lattice}.  Approximating the triclinic crystal lattice of
$\rm(TMTSF)_2X$ by the orthogonal one, the magic Lebed angles can be
written as
\begin{equation}
  \frac{B_y}{B_z}\,\frac{c}{b}=\frac{n}{m} \quad \Leftrightarrow
  \quad \sin\varphi\,\tan\theta=\frac{n}{m}\,\frac{b}{c},
\label{nm}
\end{equation}
where $n$ and $m$ are integer numbers, and $\bm
B=(B_x,B_y,B_z)=B(\sin\theta\cos\varphi,\sin\theta\sin\varphi,\cos\theta)$
is the magnetic field.  Experimentally, the Lebed effect is the most
pronounced for $m=1$.  Lee and Naughton \cite{Naughton98a,Naughton98b}
studied AMRO for generic orientations of $\bm B$, where all three
effects coexist.  They found that the Lebed oscillations are enhanced
when $B_x\neq0$ \cite{Naughton98a}, and the DKC oscillations still
exist in the presence of $B_y\neq0$ \cite{Naughton98b}.

Although initially the different types of AMRO were treated as
separate phenomena, a unified picture emerged in the recent years due
to substantial experimental and theoretical progress.  A
three-dimensional visualization of the experimentally measured
$\sigma_{zz}(\bm B)$ (Ref.~\onlinecite{Kang07a}) demonstrated that the different
types on AMRO can be viewed as modulations of the basic Lebed
resonances.  Measurements with carefully placed electric contacts
\cite{Kang07b} proved that AMRO exist only in the transverse
resistance $R_{zz}$ and not in the longitudinal resistance $R_{xx}$
along the chains.  Theory always predicted this difference, but many
experiments observed AMRO in $R_{xx}$ as well because of the mixing
between different components of the conductivity tensor.  AMRO were
found not only in the dc conductivity, but also in the ac conductivity
at microwave frequencies \cite{Ardavan98,Takahashi05}.  The ac
measurements were interpreted in terms of the so-called period orbit
resonance (POR) \cite{Hill97}, which is a generalization of the
cyclotron resonance to more complicated (e.g.,\ open) Fermi surfaces
\cite{Blundell96}.  The ac resonances occur at the angles depending on
frequency $\omega$ and deviating from Eq.~(\ref{nm})
\cite{Ardavan98,Takahashi05}, so the Lebed magic angles are not truly
magic \cite{Takahashi06,Hill08}.  This observation eliminates
theoretical scenarios proposing a radical change in the ground state
of the system depending on the magnetic-field orientation along the
magic or non-magic angles.  This conclusion is also supported by the
absence of any angular effect in NMR \cite{Wu05}.

Given these experimental facts, AMRO most likely represent some sort
of a resonance effect in the dc and ac transport coefficients.  The
first theoretical calculation along these lines was done in
Ref.~\onlinecite{Osada92} using the Kubo formula with the electron
wave functions for a magnetic field in the $(y,z)$ plane.  This
quantum-mechanical calculation was then generalized to include the
$B_x$ component of the magnetic field \cite{Lebed03} and the anion
superstructure of $\rm(TMTSF)_2ClO_4$ \cite{Ha05,Ha06}.  In another
theoretical approach, the Boltzmann kinetic equation was solved for a
constant relaxation time $\tau$ by using quasiclassical electron
trajectories on the Fermi surface
\cite{Osada96,Lebed97b,Naughton98a,Naughton98b,Hill97,Blundell96,Osada99,Yoshino99a,Yoshino99b,Kobayashi06}.
This solution can be written in a general form using the so-called
Shockley tube integral \cite{Shockley50} or the Chambers formula
\cite{Chambers52}, see also the book \cite{book-Ziman}.  In the third
theoretical approach, the interlayer conductivity was calculated using
a perturbation theory in the electron-tunneling amplitude between two
layers \cite{McKenzie,Osada03,Cooper06}.  In this approach, AMRO
originate from the Aharonov-Bohm quantum interference in interlayer
tunneling in the presence of a magnetic field \cite{Cooper06}.  All
these three seemingly different theoretical approaches produce the
same final results and are essentially equivalent.

Despite substantial progress in understanding of AMRO in Q1D
conductors, some experimental results remain unexplained.  One open
problem is the angular oscillations of the Nernst effect \cite{Wu03}.
Another unresolved problem is the angular minimum and saturation of
the interlayer resistivity $R_{zz}$ observed for a magnetic field in
the $y$ direction \cite{Chaikin92c,Kang07a,Naughton98b,Kang06}.
Although the manifestations of AMRO are qualitatively similar in all
members of the $\rm(TMTSF)_2X$ family, direct comparison of the
measurements in $\rm(TMTSF)_2PF_6$, $\rm(TMTSF)_2ClO_4$, and
$\rm(TMTSF)_2ReO_4$ shows substantial differences \cite{Kang06}.

For a magnetic-field rotation in the $(y,z)$ plane with $B_x=0$, only
three strong Lebed peaks in $\sigma_{zz}$ with $n=0,\:\pm1$ are
observed in $\rm(TMTSF)_2PF_6$ \cite{Chaikin92c,Kang06}.  When special
care is taken to ensure that $B_x=0$, the very weak peaks with
$n=\pm2$ in $\rm(TMTSF)_2PF_6$ disappear completely \cite{Kang07a}.
In contrast, in $\rm(TMTSF)_2ReO_4$, strong Lebed oscillations are
observed up to $n=\pm11$ \cite{Kang03}.  In $\rm(TMTSF)_2ClO_4$, the
Lebed oscillations are much weaker in amplitude than in
$\rm(TMTSF)_2PF_6$ and $\rm(TMTSF)_2ReO_4$ \cite{Kang06}, but many
Lebed resonance can be detected after differentiation of the data with
respect to the angle of rotation \cite{Osada91,Naughton91}.  The
strength of the DKC oscillations is also very different in these
materials.  The DKC oscillations are quite strong in
$\rm(TMTSF)_2ClO_4$, where they were originally discovered
\cite{Chaikin94a}.  In $\rm(TMTSF)_2PF_6$,
Ref.~\onlinecite{Chaikin95c} found very weak DKC oscillations, but
Ref.~\onlinecite{Kang06} found them to be substantial.  However, in
$\rm(TMTSF)_2ReO_4$, the DKC oscillations are extremely weak and
almost invisible \cite{Kang06}.  This dramatic difference in
manifestations of AMRO in the three materials requires a theoretical
explanation.

When a magnetic field is rotated in the $(y,z)$ plane at $B_x=0$, the
theoretical calculations cited above show that the Lebed peaks in
$\sigma_{zz}$ can exist only for those magic angles $(n,m)$ where the
interchain tunneling amplitudes in the directions $n\bm b+m\bm c$ are
present \cite{Osada92,Chashechkina01}.  It is reasonable to expect
that the interplane tunneling amplitudes in $\rm(TMTSF)_2PF_6$ exist
between the nearest and next-nearest chains in the $\bm c$ and $\bm
c\pm\bm b$ directions (see Fig.\ \ref{fig:lattice}).  This would
explain why only the Lebed resonance with $n=0,\:\pm1$ are observed in
$\rm(TMTSF)_2PF_6$.  However, many magic angles with big numbers $n$
are observed in $\rm(TMTSF)_2ClO_4$ and $\rm(TMTSF)_2ReO_4$.  It is
hard to imagine that direct electron overlap exists between the chains
separated by 11 interchain distances.

One way to resolve this problem is to take into account the nonlinear
electron dispersion along the chains.  (All theoretical papers cited
above make a linearized approximation for the electron dispersion
along the chains.)  The first attempt in this direction was made in
Ref.~\onlinecite{Maki92}, and a more systematic study was presented in
Refs.~\onlinecite{Lebed04a} and \onlinecite{Lebed04b}.  The nonlinearity can indeed
generate an effect similar, albeit not completely equivalent, to the
presence of many interchain tunneling amplitudes.  However, the
nonlinearity alone is not sufficient to explain the differences in
AMRO between the three compounds.  Another problem is the absence of
the DKC oscillations in $\rm(TMTSF)_2ReO_4$.  One might think that
quantum coherence is too low in this material, but the existence of 21
Lebed oscillations clearly refutes this idea \cite{Kang03}.  We see
that a detailed theoretical understanding of AMRO in the
$\rm(TMTSF)_2X$ materials is challenging and requires additional
ideas.

We believe that the key to understanding the differences in AMRO is
the presence of anion ordering in $\rm(TMTSF)_2ClO_4$ and
$\rm(TMTSF)_2ReO_4$ and its absence in $\rm(TMTSF)_2PF_6$.  $\rm PF_6$
is an octagonal centrosymmetric anion, which does not experience any
orientational ordering at low temperatures.  In contrast, $\rm ClO_4$
and $\rm ReO_4$ are tetragonal anions without inversion symmetry.
Because their crystal sites have inversion symmetry, these anions have
two different orientations of the same energy.  At low temperatures,
the anions experience orientational ordering and produce crystal
superstructures \cite{book-Ishiguro} with the wave vectors $\bm
Q=(0,1/2,0)$ in $\rm(TMTSF)_2ClO_4$ (under ambient pressure) and $\bm
Q=(0,1/2,1/2)$ in $\rm(TMTSF)_2ReO_4$ (under pressure greater than
about 10 kbar), as shown in Fig.\ \ref{fig:lattice}.  Formation of a
crystal superstructure affects electron spectrum by folding the
Brillouin zone.  In this paper, we show that reconstruction of the
electron dispersion caused by the anion ordering generates effective
tunneling amplitudes between many distant chains.  This effect
explains why many Lebed angles are observed in $\rm(TMTSF)_2ReO_4$ and
$\rm(TMTSF)_2ClO_4$, but not in $\rm(TMTSF)_2PF_6$.  It also explains
why the magic angles [Eq.~(\ref{nm})] are observed only for odd $n$ in
$\rm(TMTSF)_2ReO_4$ (Ref.~\onlinecite{Kang03}) and only for even $n$ in
$\rm(TMTSF)_2ClO_4$ (Refs.~\onlinecite{Osada91} and \onlinecite{Naughton91}) at $m=1$.  We also
explain the differences in the DKC oscillations within the same
framework.

In contrast to the previous theories of AMRO for the anion
superstructure of $\rm(TMTSF)_2ClO_4$
\cite{Ha05,Ha06,Yoshino99b,Lebed04a}, we take into account the direct
effect of anion ordering on the interlayer tunneling amplitude, which
is especially important for $\rm(TMTSF)_2ReO_4$.  In this way, we can
capture the characteristic features of AMRO in the three compounds
without invoking the nonlinearity of the longitudinal electron
dispersion \cite{Lebed04a,Lebed04b}.

In the second part of the paper (Sec.~\ref{Sec:Interband}), we study
the effect of a strong magnetic field parallel to the layers.  We show
that, when $B_y$ is strong enough and exceeds a certain threshold
related to the anion gap $E_g$, the interlayer tunneling between
different branches of the folded electron dispersion becomes possible,
and $\sigma_{zz}$ should increases sharply.  Experimental observation
of this effect would allow direct measurement of $E_g$.  This effect
can be also applied to study the interband tunneling in
$\kappa$-$\rm(ET)_2Cu(NCS)_2$.  A theory of this effect cannot be
formulated within the framework of quasiclassical orbits on a warped
Fermi surface.  We calculate an interlayer conductivity in the presence
of anion ordering using the quantum limit, where the electron wave
functions are confined to the layers due to a strong parallel magnetic
field \cite{Yakovenko87,Yakovenko88,Lebed05,Joo06}.

\section{Calculation of interlayer conductivity}
\label{Sec:Conductivity}

The general form of the electron dispersion in a Q1D metal is
\begin{equation}
  \varepsilon(\bm k) = \pm\hbar v_F(k_x \mp k_F) +
  \varepsilon_\perp(k_y,k_z),
\label{energy}
\end{equation}
where the energy $\varepsilon$ is measured from the Fermi energy, and
$\bm k=(k_x,k_y,k_z)$ is the electron wave vector.  Here we linearize
the dispersion along the chains with the Fermi velocity $v_F$ near the
Fermi wave vectors $\pm k_F$.  There are two sheets of the open Fermi
surface, but we present calculations only for the sheet with $+v_F$.
Since $t_c \ll t_b$, we can expand the transverse dispersion
$\varepsilon_\perp$ to the lowest order in the interlayer tunneling
amplitude $t_c$,
\begin{equation}
  \varepsilon_\perp(k_y,k_z)=2t_b\varepsilon_y(k_yb)
  +2t_cf(k_yb)\cos(k_zc).
\label{f}
\end{equation}
For a simple model with electron tunneling between the nearest chains
in the absence of a superstructure, Eq.~(\ref{f}) reduces to a
standard tight-binding expression with $\varepsilon_y(k_y)=\cos(k_yb)$
and $f(k_yb)=1$.  However, we will show in Secs.\ \ref{Sec:ReO4} and
\ref{Sec:ClO4} that a nontrivial function $f(k_yb)$ appears in the
interlayer tunneling term in the presence of anion ordering.  This
effect was not considered in previous literature and plays a crucial
role in our consideration.

From the dispersion relation (\ref{energy}), we obtain the electron
velocity $\bm v = \partial\varepsilon/\hbar\partial\bm k$,
\begin{equation}
  v_x=v_F, \quad
  v_y \approx \frac{2t_b}{\hbar} \frac{d\varepsilon_y}{dk_y},\quad
  v_z=-\frac{2t_cc}{\hbar}f(k_yb) \sin(k_zc).
\label{vz}
\end{equation}
In the quasiclassical approximation, the time-dependent electron wave
vector $\bm k^{(t)}$ follows the equation of motion,
\begin{equation}
  \hbar\,\frac{d\bm k^{(t)}}{dt}=e\bm v^{(t)}\times\bm B,
\end{equation}
where $e$ is the electron charge, and the magnetic field $\bm B$ is in
the SI units.  Given that $v_x = v_F \gg v_z $, we find
\begin{eqnarray}
  \frac{dk_y^{(t)}}{dt} \approx -\frac{e v_F B_z}{\hbar}, \;
  k_y^{(t)} = -\frac{\omega_ct}{b} + k_y^{(0)}, \;
  \omega_c=\frac{ebv_FB_z}{\hbar},
\label{ky}
\end{eqnarray}
where $\omega_c$ is the analog of the cyclotron frequency for the open
Fermi surface. The equation of motion for $k_z$ is
\begin{equation}
  dk_z = \frac{e}{\hbar}\left( v_FB_ydt
  - \frac{2t_bB_x}{\hbar} \frac{d\varepsilon_y}{dk_y}\, dt \right).
\end{equation}
Using $dk_y/dt$ from Eq.\ (\ref{ky}), we get
\begin{equation}
  ck_z^{(t)}= B_y'\omega_ct+ B_x'\varepsilon_y(k_y^{(t)})+ck_z^{(0)},
\label{kz}
\end{equation}
where we introduced the dimensionless parameters
\begin{eqnarray}
  B_y'=\frac{B_y}{B_z}\frac{c}{b}, \quad
  B_x'=\frac{B_x}{B_z}\frac{2t_bc}{\hbar v_F}.
\label{Bxy'}
\end{eqnarray}
The variables $B_y'$ and $B_x'$ are proportional to the tangents of
the magnetic field projections onto the $(y,z)$ and $(x,z)$ planes,
respectively.

The interlayer conductivity $\sigma_{zz}$ is given by the Shockley tube
integral \cite{book-Ziman},
\begin{equation}
  \sigma_{zz} =
  \frac{4e^2}{\hbar}
  \int\!\!\!\!\int\frac{dk_y^{(0)}dk_z^{(0)}}{(2\pi)^3v_F}
  \int\limits_{-\infty}^0 dt\,
  v_z(\bm k^{(0)})v_z(\bm k^{(t)})e^{t(1/\tau-i\omega)},
\label{corr}
\end{equation}
where $\tau$ is a relaxation time, and the factor 4 comes from the two
spin projections and the two sheets of the Fermi surface.
Substituting Eqs.\ (\ref{vz}),\ (\ref{ky}), and (\ref{kz}) into Eq.\
(\ref{corr}), we find the real part of $\sigma_{zz}$
\begin{eqnarray}
  \!\!\!\!\!\!
  && \sigma_{zz} = \frac{e^2t_c^2c}{\pi^2\hbar^3\omega_cv_Fb}{\mathcal Re}
  \sum_{\mp} \int\limits_0^{2\pi} d\phi \int\limits_0^\infty d\eta\,
  f(\phi)f(\phi+\eta) 
\label{integral} \\ 
   \!\!\!\!\!\! && \times
  \exp\{iB_{x}'[\varepsilon_y(\phi)-\varepsilon_y(\phi+\eta)]
  -\eta[1/\omega_c\tau-iB_y'\mp i\omega/\omega_c]\},
\nonumber
\end{eqnarray}
where $\phi=bk_y^{(0)}$ and $\eta=-\omega_ct$. Expanding the periodic
functions $f(\phi)\,e^{iB_x'\varepsilon(\phi)}$ in Eq.\
(\ref{integral}) into the Fourier series with the coefficients
\begin{equation}
  A_n(B_x')=\frac{1}{2\pi} \int\limits_0^{2\pi}
  e^{-in\phi}f(\phi)e^{iB_{x}'\varepsilon_y(\phi)}\,d\phi,
\label{An}
\end{equation}
we obtain
\begin{equation}
  \frac{\sigma_{zz}}{\sigma_0}= \frac{1}{2}
  \sum_{\mp}\sum_{n=-\infty}^\infty \frac{|A_n(B_{x}')|^2}
  {1+(\omega_c\tau)^2(n-B_{y}'\mp\omega/\omega_c)^2}.
\label{sigma}
\end{equation}
Here $\sigma_0=(4e^2t_c^2\tau c)/(\pi\hbar^3v_Fb)$ is the interlayer
dc conductivity at $\bm B=0$, and the $\pm$ terms are the
contributions from the two sheets of the Fermi surface.  In the rest
of the paper, we shall focus on the dc conductivity $\sigma_{zz}$ at
$\omega=0$, although Eq.~(\ref{sigma}) also gives the ac conductivity.
  
The Lebed effect corresponds to the resonant peaks of $\sigma_{zz}$ in
Eq.~(\ref{sigma}) achieved at $B_y'=n$, where the condition (\ref{nm})
for $m=1$ is satisfied.  In a simple model without anion ordering,
where $\varepsilon_y=\cos(k_yb)$ and $f=1$, Eq.\ (\ref{An}) reduces to
$A_n(B_x')=i^nJ_n(B_x')$, where $J_n$ is the Bessel function.  In this
case, Eq.\ (\ref{sigma}) reproduces the result found in Refs.\
\onlinecite{Cooper06,McKenzie,Osada03,Lebed03,Kobayashi06}.  However,
the coefficients $J_n(B_x')$ vanish for $n\neq0$ at $B_x=0$, so there
are no Lebed oscillations in this model for a magnetic-field rotation
in the $(y,z)$ plane.  The DKC effect originates from the oscillations
of $J_n(B_x')$ vs $B_x'$ in the numerator of Eq.~(\ref{sigma}).

Interestingly, Eq.~(\ref{sigma}) with $|A_n|^2=J_n^2(B_x')$ and
$\omega=0$ is exactly the same as the equation \cite{Oliver05,Berns06}
that describes the Mach-Zehnder interference in a superconducting
qubit driven by an ac electric field and subjected to a dc bias
\cite{Oliver05,Berns06,Sillanpaa06,Izmalkov08,Ashhab07}.  The two
states of the qubit correspond to the two adjacent layers of a Q1D
conductor coupled by the tunneling amplitude $t_c$.  The frequency of
the ac field for the qubit maps to the frequency $\omega_c$ in
Eq.~(\ref{ky}), the detuning of the qubit maps to
$B_y'\omega_c=ecv_FB_y/\hbar$, and the amplitude of the ac modulation
maps to $B_x'$ in Eq.~(\ref{Bxy'}).  The contour plot of
Eq.~(\ref{sigma}) shown in Fig.~2 of Ref.~\onlinecite{Cooper06} is
exactly the same as in Refs.~\onlinecite{Oliver05} and \onlinecite{Berns06}, and it
represents the so-called Bessel staircase.  The same equation also
appears in the theory of laser cooling in ion traps \cite{DeVoe89}.
This correspondence is not just a mathematical curiosity, but it also reflects
profound similarity between these highly coherent quantum system,
where the oscillatory patterns are caused by phase interference due to
applied electric and magnetic fields.

\section{Interlayer conductivity in \boldmath $\rm(TMTSF)_2PF_6$ 
without anion ordering}
\label{Sec:PF6}

Let us first discuss the case of $\rm(TMTSF)_2PF_6$, which does not
have anion ordering.  In order to observe more than one Lebed angle,
we need to introduce the tunneling amplitude $t_c'$ between
next-nearest neighboring chains, as shown in Fig.\
\ref{fig:lattice}(b).  Including this term in the transverse
dispersion (\ref{f}), we find for $\rm(TMTSF)_2PF_6$
\begin{equation}
  \varepsilon_y(\phi)=\cos\phi, \quad
  f(\phi)=1 + 2\frac{t_c'}{t_c}\cos\phi, \quad
  \phi=bk_y.
\label{PF6}
\end{equation}
In a more general case, where the amplitudes $t_n$ corresponding to
the tunneling vectors $\bm c+n\bm b$ are present, the transverse
dispersion relation can be written as
\begin{equation}
  \varepsilon_\perp(k_y,k_z)=2t_b\cos(k_yb)
  +2\sum_l t_l\cos(k_zc+lk_yb).
\label{t_m}
\end{equation}
Equation (\ref{PF6}) is the special case of Eq.~(\ref{t_m}) with $t_0=t_c$
and $t_{\pm1}=t_c'$.

Generalizing the derivation presented in Sec.~\ref{Sec:Conductivity}
to the transverse dispersion relation (\ref{t_m}), we find that the
interlayer conductivity $\sigma_{zz}$ is given by Eq.~(\ref{sigma})
with the following coefficients $A_n$ (Ref.~\onlinecite{error})
\begin{equation}
  A_n(B_x') = \frac{1}{t_c}\sum_l i^{n+l} t_l J_{n+l}(B_x').
\label{Anl}
\end{equation}
In the case of $\rm(TMTSF)_2PF_6$, Eqs.~(\ref{An}) and (\ref{PF6}) or
Eq.~(\ref{Anl}) give
\begin{equation}
  A_n(B_x') = i^nJ_n(B_x') + i^{n+1}\frac{t_c'}{t_c}J_{n+1}(B_x')
  + i^{n-l}\frac{t_c'}{t_c}J_{n-1}(B_x').
\label{An+-1}
\end{equation}
Substituting Eq.~(\ref{An+-1}) into Eq.~(\ref{sigma}), we obtain
$\sigma_{zz}$ for $\rm(TMTSF)_2PF_6$.  When $B_x'=0$,
Eq.~(\ref{An+-1}) gives non-zero coefficients $A_n$ only for $n=0$ and
$n=\pm1$.  Thus, Eq.~(\ref{sigma}) exhibits the Lebed peaks only at
$n=0$ and $n=\pm1$ with the heights proportional to $t_c^2$ and
$(t_c')^2$ for a magnetic-field rotation in the $(y,z)$ plane.

When we consider the DKC oscillations at $B_y'=0$, i.e.,\ for a
magnetic-field rotation in the $(x,z)$ plane, the sum in
Eq.~(\ref{sigma}) is dominated by the term with $n=0$, because the
other terms have the big factor $(\omega_c\tau)^2$ in the denominator.
Keeping only the term with $n=0$ and using Eq.~(\ref{An+-1}), we can
write approximately,
\begin{equation}
  \frac{\sigma_{zz}(B_x')}{\sigma_0}  \approx
  \left|J_0(B_x') + 2i\frac{t_c'}{t_c}J_1(B_x')\right|^2.
\label{sigma-n0}
\end{equation}
When $t_c'=0$, Eq.~(\ref{sigma-n0}) vanishes for the angles where
$J_0(B_x')=0$, which is a manifestation of the DKC oscillations.
However, in the presence of $t_c'\neq0$, Eq.~(\ref{sigma-n0}) does not
vanish for any angles, so the DKC oscillations are partially
suppressed, although some modulation of $\sigma_{zz}$ vs.\ $B_x'$
remains.  We see that the presence of tunneling amplitudes $t_l$ to
more distant chains enhances the Lebed oscillations but suppresses
the DKC oscillations.  This conclusion was already made in
Ref.~\onlinecite{Cooper06}.

\section{Anion ordering in \boldmath $\rm(TMTSF)_2ReO_4$}
\label{Sec:ReO4}

\begin{figure}
\includegraphics[width=0.9\linewidth]{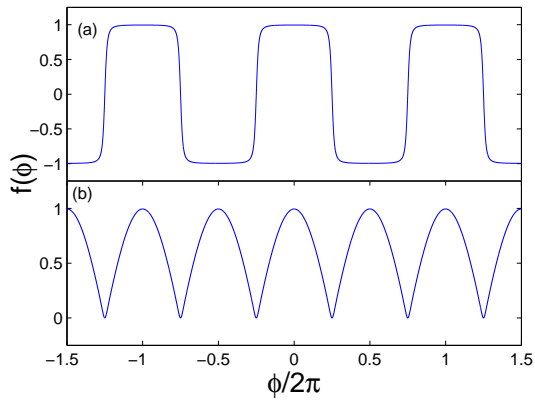}
\caption{(Color online) (a) Plot of the function $f(\phi)$ given by
  Eq.~(\ref{f1}). (b) Plot of the second term in $f(\phi)$ given by
  Eq.~(\ref{f2}).  In both plots, $E_g/2t_b = 0.1$.}
\label{fig:fphi}
\end{figure}

The $\rm ReO_4$ anions order with the wave vector $\bm Q =
(0,1/2,1/2)$ under pressure.  This causes the energies of the odd and
even chains to split by $\pm E_g$, as illustrated in Fig.\
\ref{fig:lattice}(a).  The Hamiltonian of interchain tunneling is
described by a $2\times2$ matrix representing the even and odd chains
\cite{t_c'}:
\begin{equation}
  H_\perp = \left( \begin{array}{cc}
  E_g & 2t_b\cos(k_yb)+ 2t_c\cos(k_zc) \\
  \rm c.c. & -E_g
  \end{array} \right).
\label{H1}
\end{equation}
The eigenvalues of the matrix (\ref{H1}) give the transverse electron
dispersion relation,
\begin{equation}
  \varepsilon_\perp=\pm\sqrt{[2t_b\cos(k_yb)+2t_c\cos(k_zc)]^2+E_g^2}.
\label{sqrt1}
\end{equation}
Expanding Eq.~(\ref{sqrt1}) to the zeroth and first order in $t_c$, we
find the functions $\varepsilon_y(k_y)$ and $f(k_y)$ in Eq.\ (\ref{f})
\begin{eqnarray}
  \varepsilon_y(\phi) &=& \pm\sqrt{\cos^2\phi+(E_g/2t_b)^2},
  \quad \phi=bk_y,
\label{Ey1} \\
  f(\phi) &=& \pm\frac{\cos\phi}{\sqrt{\cos^2\phi+(E_g/2t_b)^2}}.
\label{f1}
\end{eqnarray}
The function $f(\phi)$ in Eq.\ (\ref{f1}) is close to a square wave
for $E_g/t_b\ll1$, as shown in Fig.\ \ref{fig:fphi}(a).  Its Fourier
coefficients $A_n$, given by Eq.~(\ref{An}) with $B_x'=0$, are
non-zero only for odd $n$ and decay as $1/n$.  Transforming
Eq.~(\ref{f}) from the momentum space to the real space, we find that
the Fourier coefficients of $f(k_yb)$ generate effective interplane
tunneling amplitudes along the vectors $\bm c+n\bm b$ with odd $n$,
which are shown in Fig.\ \ref{fig:lattice}(a) by the arrows.
Initially, the model has only the tunneling amplitudes $t_b$ and $t_c$
between the nearest chains, but the anion ordering generates effective
tunneling amplitudes between many chains.  The higher-order expansion
of Eq.\ (\ref{sqrt1}) in $t_c$ would generate effective tunneling
amplitudes along the vectors $m\bm c+n\bm b$ with $m$ and $n$ of the
same parity between the sites of the same type, either open circles or
closed circles in Fig.\ \ref{fig:lattice}(a).  However, one should
keep in mind that this heuristic real-space picture \cite{Kang03} is
an oversimplification, and an accurate calculation in the momentum
space should be performed as described above.

\begin{figure}
\includegraphics[width=0.9\linewidth]{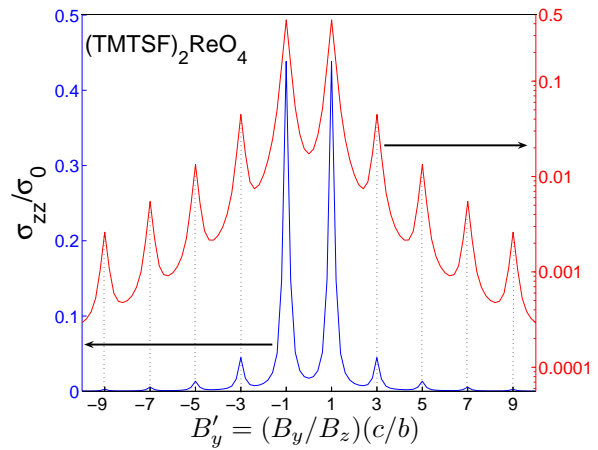}
\caption{(Color online) Normalized interlayer conductivity
  $\sigma_{zz}/\sigma_0$ calculated from Eq.~(\ref{sigma}) for
  $\rm(TMTSF)_2ReO_4$ and plotted vs.\ $B_y'$ at $B_x'=0$, shown in
  the linear (left) and logarithmic (right) scales.}
\label{fig:ReO4}
\end{figure}

In Fig.\ \ref{fig:ReO4} we show the normalized dc conductivity
calculated from Eq.\ (\ref{sigma}) for $B_x'=0$ and
$\omega_c\tau=\sqrt{50}$ using the Fourier coefficients $A_n$ from
Eq.~(\ref{An}).  Since $A_n\ne0$ only for odd $n$, therefore
$\sigma_{zz}$ has peaks only at the odd Lebed angles, as shown in
Fig.\ \ref{fig:ReO4} and observed in $\rm(TMTSF)_2ReO_4$
\cite{Kang03}.  The higher-order expansion of Eq.\ (\ref{sqrt1}) in
$t_c$ would generate peaks at the Lebed magic angles with $m$ and $n$
of the same parity in Eq.\ (\ref{nm}), as observed in Ref.\
\onlinecite{Kang03}.  Because of the anion superstructure,
Eq.~(\ref{sqrt1}) is highly non-linear in $\cos\phi$; so its Fourier
expansion generates a big number of harmonics, which produce a big
number of Lebed peaks in AMRO.  This is the qualitative reason why so
many Lebed peaks are observed in $\rm(TMTSF)_2ReO_4$, in contrast to
$\rm(TMTSF)_2PF_6$, which has no anion superstructure.

\begin{figure}
\includegraphics[width=0.9\linewidth]{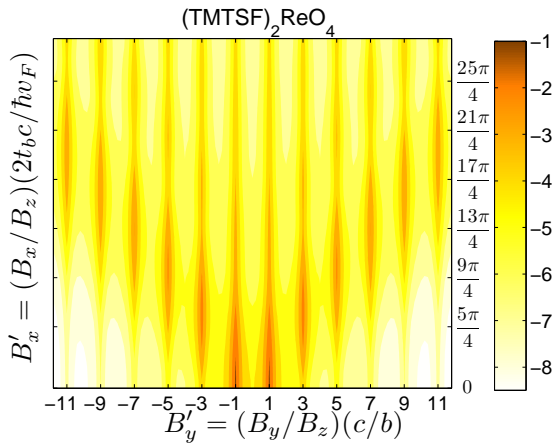}
\caption{(Color online) Contour plot of $\ln(\sigma_{zz}/\sigma_0)$
  calculated from Eq.\ (\ref{sigma}) for $\rm(TMTSF)_2ReO_4$ at
  $\omega_c\tau = \sqrt{50}$.}
\label{fig:contReO4}
\end{figure}

Figure~\ref{fig:contReO4} shows a contour plot of
$\ln(\sigma_{zz}/\sigma_0)$ vs.\ $B_x'$ and $B_y'$, as calculated from
Eq.\ (\ref{sigma}) using Eqs.\ (\ref{An}), (\ref{Ey1}), and
(\ref{f1}).  The conductivity is maximal at the vertical lines
corresponding to the odd Lebed magic angles.  At a fixed Lebed angle,
the weak modulation of $\sigma_{zz}$ vs.\ $B_x'$ (along a vertical
line) corresponds to the DKC oscillations.  Figure \ref{fig:contReO4}
shows that the DKC oscillations are very weak, because the
coefficients $A_n(B_x')$ [Eq.~(\ref{An})] do not have zeros vs.\ $B_x'$ in
the presence of anion ordering, unlike the Bessel functions
$J_n(B_x')$ in a simple model.  This is a theoretical explanation of
why the DKC oscillations in $\rm(TMTSF)_2ReO_4$ are very weak and
barely detectable experimentally \cite{Kang03}.

\section{Anion ordering in \boldmath $\rm(TMTSF)_2ClO_4$}
\label{Sec:ClO4}

In the case of $\rm(TMTSF)_2ClO_4$, in order to observe multiple Lebed
angles, we need to take into account the tunneling amplitude $t_c'$
introduced in Sec.~\ref{Sec:PF6} and shown in Fig.\
\ref{fig:lattice}(b). For the anion ordering with $\bm Q=(0,1/2,0)$,
the interchain tunneling is described by the Hamiltonian,
\begin{equation}
  H_\perp = \left( \begin{array}{cc}
  E_g + 2t_c \cos(k_zc) & \cos(k_y b)[2t_b+ 4t_{c}'\cos(k_zc)] \\
  \rm c.c. & -E_g+2t_c \cos(k_zc)
  \end{array} \right).
  \label{H2}
\end{equation}
The eigenvalues of the matrix (\ref{H2}) give the transverse electron
dispersion relation,
\begin{equation}
  \varepsilon_\perp = 2t_c \cos(k_zc) \pm \sqrt{\cos^2 (k_yb)[2t_b+
  4t_{c}'\cos(k_zc)]^2+E_g^2}.
\label{disprel}
\end{equation}
Expanding Eq.\ (\ref{disprel}) to the zeroth and first order in $t_c$
and comparing it with Eq.\ (\ref{f}), we find $\varepsilon_y(\phi)$ to be
the same as in Eq.\ (\ref{Ey1}) and
\begin{equation}
  f(\phi)=1 \pm \frac{t_c'}{t_{c}}\frac{2\cos^2\phi}
  {\sqrt{\cos^2\phi+(E_g/2t_b)^2}}.
\label{f2}
\end{equation}

\begin{figure}
\includegraphics[width=0.9\linewidth]{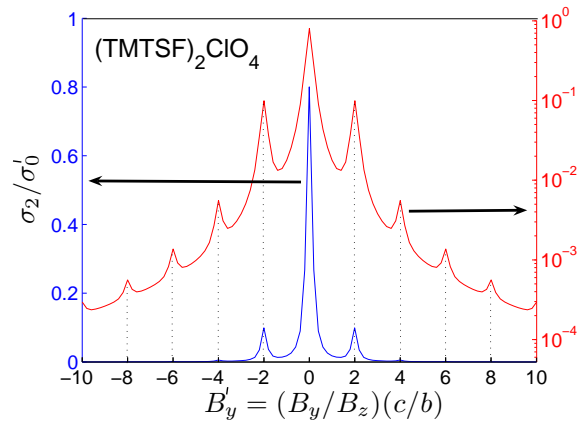}
\caption{(Color online) Plot of $\sigma_2/\sigma_0'$ vs.\ $B_y'$ at
$B_x'=0$, shown in the linear (left) and logarithmic (right) scales.
$\sigma_2$ is the contribution to $\sigma_{zz}$ in Eq.\ (\ref{sigma})
from the second term in Eq.\ (\ref{f2}), and
$\sigma_0'=\sigma_0(t_c'/t_c)^2$.}
\label{fig:ClO4}
\end{figure}

Only the second term in Eq.\ (\ref{f2}) generates the coefficients
$A_n$ with $n\neq0$ when substituted into Eq.~(\ref{An}) at $B_x=0$.
For $E_g/t_b\ll1$, this term is close to a rectified cosine signal, as
shown in Fig.\ \ref{fig:fphi}(b), and its Fourier coefficients decay
as $1/n^2$ for large $n$.  It has non-zero Fourier coefficients only
for even $n$, thus $\sigma_{zz}$ vs $B_y'$ has peaks at the even
Lebed angles, as shown in Fig.\ \ref{fig:ClO4} for $B_x' = 0$ and
observed experimentally in $\rm(TMTSF)_2ClO_4$
\cite{Osada91,Naughton91}.  Because the second term in Eq.~(\ref{f2})
is highly nonlinear in $\cos\phi$, it generates many harmonics and
many Lebed peaks.  However, they decay with the increase in $n$ faster
in $\rm(TMTSF)_2ClO_4$ than in $\rm(TMTSF)_2ReO_4$.  Moreover, because
$t_c'$ is small, the Lebed oscillations in $\rm(TMTSF)_2ClO_4$ are
weak, in agreement with the observations in Refs.\
\onlinecite{Osada91,Naughton91,Kang06}.  As discussed in
Sec.~\ref{Sec:PF6}, the DKC oscillations are controlled by the
coefficient $A_0(B_x')$ in Eq.~(\ref{sigma}).  The first term in Eq.\
(\ref{f2}) gives the main contribution to $A_0(B_x')$, proportional to
$J_0(B_x')$.  Thus, the DKC oscillations are relatively strong in
$\rm(TMTSF)_2ClO_4$, as observed in Refs.\
\onlinecite{Chaikin94a,Kang06}, although they are somewhat reduced by
the second term in Eq.\ (\ref{f2}).

We conclude that the different types of anion ordering in
$\rm(TMTSF)_2ReO_4$ and $\rm(TMTSF)_2ClO_4$ can indeed explain the
characteristic features of AMRO in these materials.  In
$\rm(TMTSF)_2ReO_4$, the Lebed oscillations are strong and numerous,
but the DKC oscillations are very weak.  In $\rm(TMTSF)_2ClO_4$, the
Lebed oscillations are numerous, but weak, whereas the DKC
oscillations are relatively strong.  On the other hand, there is no
anion superstructure in $\rm(TMTSF)_2PF_6$.  This material exhibits a
few but strong Lebed oscillations and partially suppressed DKC
oscillations.

\section{Interband tunneling in a strong magnetic field parallel to 
the layers}
\label{Sec:Interband}

\begin{figure}
\includegraphics[width=0.9\linewidth]{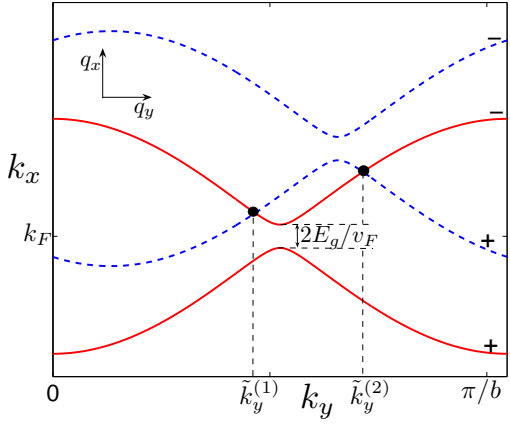}
\caption{(Color online) Fermi surfaces of two adjacent layers shifted
  by the vector $\bm q$ of Eq.~(\ref{q}) due to an in-plane magnetic
  field. The Fermi surfaces for each layer (the solid lines and the
  dashed lines) consist of two bands separated by the gap $2E_g/v_F$
  due to anion ordering and labeled $+$ and $-$.}
\label{fig:double}
\end{figure}

Folding of the Brillouin zone due to anion ordering produces two
branches (or two bands) of the electron dispersion, which we label by
the index $\alpha=\pm$ according to the sign in Eq.~(\ref{Ey1}).  The
Fermi surfaces of the two bands, obtained from Eq.~(\ref{energy}), are
shown by the two solid lines in Fig.\ \ref{fig:double} for
$E_g/t_b=0.1$. (This picture is for the Fermi-surface sheets near
$+k_F$.)

In this section, we study the interlayer conductivity in a strong magnetic
field $(B_x,B_y,0)$ parallel to the layers.  We use the formalism
developed in Refs.~\onlinecite{McKenzie,Osada03,Cooper06} and
calculate $\sigma_{zz}$ between just two layers, i.e.,\ for a bilayer.
Assuming that $t_c$ is very weak, one can argue that, in the lowest
order in $t_c$, the interlayer conductivity of a bulk multilayer
crystal is determined by the interlayer conductivity between a pair of
layers \cite{t_c'}.

The tunneling Hamiltonian between layers 1 and 2 is
\begin{eqnarray}
  && \hat{H}_c = t_c \int \hat\psi_2^\dag({\bm r})\,
  \hat\psi_1({\bm r})\, e^{i\chi({\bm r})} d^2r\,
  + \mbox{H.c.},
\label{H_perp} \\
  && \chi({\bm r})=\frac{ec}{\hbar}A_z({\bm r}), \quad
  A_z({\bm r})=B_x y- B_y x,
\label{phi}
\end{eqnarray}
where $\hat\psi_{1,2}$ are the electron destruction operators in
layers 1 and 2.  Here $A_z$ is the vector potential of the in-plane
magnetic field, and $\chi(r)$ is the corresponding gauge phase
accumulated in the process of tunneling across the interlayer spacing
$c$.  Substituting Eq.~(\ref{phi}) into Eq.~(\ref{H_perp}) and using
momentum representation in the $(x,y)$ plane, we observe that the
in-plane wave vector of the electron changes from $\bm k$ to $\bm
k+\bm q$ in the process of tunneling \cite{Cooper06}, where the vector
$\bm q$ is
\begin{equation}
  \bm q=(q_x,q_y)=\frac{ec}{\hbar}\,(B_y,-B_x).
\label{q}
\end{equation}
Thus, the Fermi surfaces of the second layer are shifted by the vector
$\bm q$ relative to the Fermi surfaces of the first layer as shown by
the two dashed lines in Fig.\ \ref{fig:double}.  A similar picture was
discussed for closed Fermi surfaces in semiconducting bilayers in
Refs.~\onlinecite{McKenzie,Eisenstein91,Simmons93,Yakovenko06}.

The interlayer conductivity $\sigma_{zz}^{\alpha\beta}$ between the
bands $\alpha$ and $\beta$ is given by the following expression
\cite{McKenzie,Mahan}
\begin{equation}
  \sigma_{zz}^{\alpha\beta} = \frac{e^2t_c^2c}{\hbar\pi} \sum_{\bm k}
  |M_{\alpha\beta}|^2 S(\bm k, E_F)S(\bm k+\bm q,E_F),
\label{sigma_zz}
\end{equation}
where $M_{\alpha\beta}=\langle\psi_\alpha^{(2)}(\bm k+\bm
q)|\psi_\beta^{(1)}(\bm k)\rangle$ is the scalar product between the
in-plane electron wave functions belonging to adjacent layers.  These
matrix elements are discussed in more detail in Appendix
\ref{App:MatrixElements}.  The total interlayer conductivity is the
sum over all bands
$\sigma_{zz}=\sum_{\alpha\beta}\sigma_{zz}^{\alpha\beta}$.  The
function $S(\bm k, E_F)$ is the spectral density of the in-plane
electron Green's function evaluated at the Fermi energy $E_F$ as a
function of the wave vector $\bm k$ \cite{McKenzie,Mahan}
\begin{equation}
  S(\bm k,E_F) = \frac{2\Gamma}{[E_F-\varepsilon(\bm k)]^2+\Gamma^2},
\label{S}
\end{equation}
where $\Gamma = \hbar/2\tau$ is the relaxation rate, and
$\varepsilon(\bm k)$ is the electron dispersion within the layer.

\begin{figure}
\includegraphics[width=0.9\linewidth]{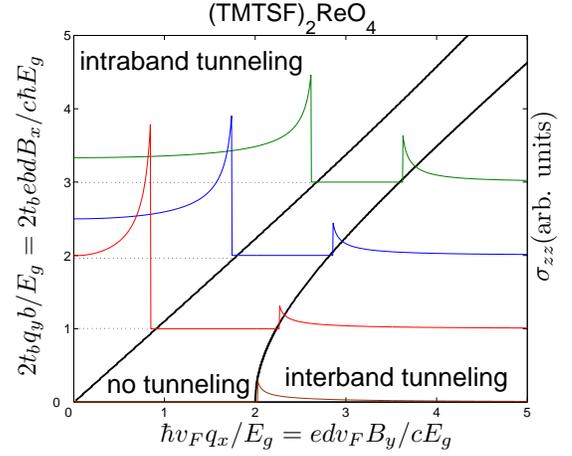}
\caption{(Color online) Phase diagram of interlayer tunneling vs.\ the
normalized in-plane magnetic-field components $B_y$ and $B_x$.
Tunneling between the same and different types of bands is possible in
the upper left and the lower right regions of the diagram,
correspondingly, and not possible in the intermediate region.  The
thin curves show the interlayer conductivity $\sigma_{zz}$ calculated
using Eq.~(\ref{dg/dk}) as a function of $B_y$ for several values of
$B_x$ for the superstructure of $\rm(TMTSF)_2ReO_4$.}
\label{fig:boundaries-ReO4}
\end{figure}

When $\Gamma$ is small, i.e.,\ when the electron quasiparticles have a
long lifetime $\tau$, the spectral function (\ref{S}) can be
replaced by a delta function, $S(\bm
k,E_F)\approx2\pi\delta[E_F-\varepsilon(\bm k)]$.  Substituting this
expression into Eq.~(\ref{sigma_zz}), we find
\begin{eqnarray}
  \sigma_{zz}^{\alpha\beta} &=& 
  \frac{e^2t_c^2c|\tilde M_{\alpha\beta}|^2}{\hbar\pi}
  \int\!\!\!\!\int dk_y\,dk_x\, 
  \delta[\hbar v_Fk_x + \alpha2t_b\varepsilon_y(k_yb)] 
\nonumber \\ 
  &\times& \delta[\hbar v_F(k_x-q_x)+\beta2t_b\varepsilon_y(k_yb-q_yb)],
\label{sigmac}
\end{eqnarray}
where the matrix element $\tilde M_{\alpha\beta}$ is evaluated at the
points where both delta functions are satisfied.  Integrating
Eq.~(\ref{sigmac}) over $k_x$, we find
\begin{equation}
  \sigma_{zz}^{\alpha\beta} = 
  \frac{e^2t_c^2c|\tilde M_{\alpha\beta}|^2}{\pi\hbar^2v_F}
  \int dk_y\, \delta[g_{\alpha\beta}(k_y)],
\label{sigmac1}
\end{equation}
where the function $g_{\alpha\beta}(k_y)$ is
\begin{equation}
  g_{\alpha\beta}(k_y)= v_Fq_x + 2t_b[\alpha\varepsilon_y(k_yb) -
 \beta\varepsilon_y(k_yb-q_yb)].
\label{g}
\end{equation}
Taking the integral (\ref{sigmac1}), we find
\begin{equation}
  \sigma_{zz}^{\alpha\beta} = 
  \frac{e^2t_c^2c}{\pi\hbar^2v_F}\sum_{\tilde k_y}
  \frac{|M_{\alpha\beta}(\tilde k_y)|^2}
  {|\partial g_{\alpha\beta}/\partial k_y|},
\label{dg/dk}
\end{equation}
where the sum is taken over the points $\tilde k_y$ where the equation
$g_{\alpha\beta}(\tilde k_y)=0$ is satisfied.  Notice that the
relaxation time $\tau$ drops out from Eq.~(\ref{dg/dk}), so
$\sigma_{zz}$ should be temperature-independent in a strong parallel
magnetic field \cite{McKenzie}.

\begin{figure}
\includegraphics[width=0.9\linewidth]{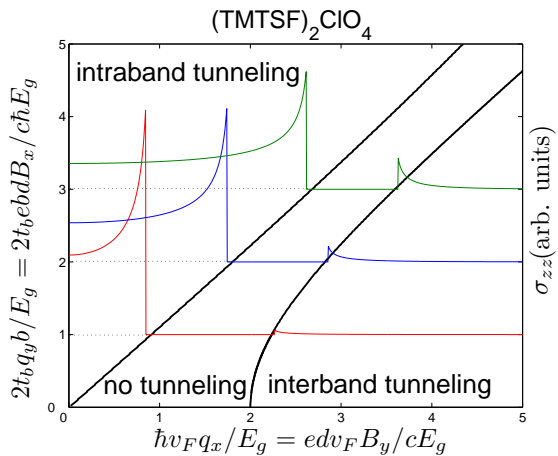}
\caption{(Color online) The same as in Fig.~\ref{fig:boundaries-ReO4},
  but for the superstructure of $\rm(TMTSF)_2ClO_4$.}
\label{fig:boundaries-ClO4}
\end{figure}

Equation (\ref{sigmac}) shows that a non-zero contribution to interlayer
conductivity comes from the points where both delta functions are
satisfied, i.e.,\ the initial and final states belong to the Fermi
surfaces of different layers.  Geometrically, these are the
intersection points $\tilde k_y^{(1)}$ and $\tilde k_y^{(2)}$ of the
solid and dashed lines in Fig.~\ref{fig:double}.  Depending on which
Fermi surfaces intersect in Fig.~\ref{fig:double}, electrons can
tunnel between different bands $\alpha,\beta=\pm$ in the folded
Brillouin zone.  The equation $g_{\alpha\beta}(\tilde k_y)=0$ has
solutions only in some regions of the $(q_x,q_y)$ space, as shown by
the thick solid lines in Figs.\ \ref{fig:boundaries-ReO4} and
\ref{fig:boundaries-ClO4}.  Above the diagonal line in Figs.\
\ref{fig:boundaries-ReO4} and \ref{fig:boundaries-ClO4}, the
interlayer tunneling is possible only between the bands of the same
type $\alpha=\beta$.  If $q_x$ exceeds the threshold value,
\begin{equation}
  \hbar v_Fq_x=B_yecv_F\geq2E_g,
\label{Delta}
\end{equation}
the interlayer tunneling between different bands, $\alpha=-\beta$,
becomes possible in the lower right region in Figs.\
\ref{fig:boundaries-ReO4} and \ref{fig:boundaries-ClO4}.  No
interlayer tunneling is possible in the intermediate region in Figs.\
\ref{fig:boundaries-ReO4} and \ref{fig:boundaries-ClO4}, where the
shifted Fermi surfaces in Fig.\ \ref{fig:double} do not cross.  The
boundaries of the regions are determined by the condition that the
displaced Fermi surface touches the other one.

The plots of the interlayer conductivity $\sigma_{zz}$, calculated
from Eq.\ (\ref{dg/dk}), are shown in Figs.\ \ref{fig:boundaries-ReO4}
and \ref{fig:boundaries-ClO4} as functions of $B_y\propto q_x$ for
several fixed values of $B_x\propto q_y$.  We observe that the
interlayer conductivity vanishes in the intermediate region and has
peaks at the boundaries.  The peaks originate from the increase in the
phase volume in the integral (\ref{sigmac}) when the two Fermi
surfaces touch each other.  Figure~\ref{fig:boundaries-ReO4} corresponds
to the anion superstructure of $\rm(TMTSF)_2ReO_4$.  We observe that,
when the magnetic field is applied along the $y$ axis ($B_x=0$), the
interlayer conductivity $\sigma_{zz}(B_y)$ is strongly suppressed
until $B_y$ exceeds the threshold, and then $\sigma_{zz}$ increases
sharply.  The value of $E_g$ can be determined from the measured
threshold field $B_y$ via Eq.~(\ref{Delta}).
Figure~\ref{fig:boundaries-ClO4} corresponds to the anion superstructure
of $\rm(TMTSF)_2ClO_4$.  In this case, the eigenfunctions of different
bands $\alpha=-\beta$ are orthogonal, so the matrix element $M_{-+}$
vanishes for $q_y=0$ (see Appendix \ref{App:MatrixElements}).  Thus,
in order to get a nonzero interlayer conductivity in
$\rm(TMTSF)_2ClO_4$, it is necessary to have a non-zero component
$B_x\neq0$, so that $q_y\neq0$.

According to the measurements in Ref.\ \onlinecite{Takahashi05}, the
Fermi velocity in $\rm(TMTSF)_2ClO_4$ is $v_F\approx10^5$ m/s.
Substituting this value and the interlayer distance $c=1.35$ nm
(Ref.~\onlinecite{book-Ishiguro}) into Eq.\ (\ref{Delta}) and using the maximal
stationary field of $45$ T available at NHMFL in Tallahassee, we find
the maximal anion gap $2E_g\approx70$ K that can be probed using this
method.  Various estimates of $E_g$ are reviewed in Ref.\
\onlinecite{Haddad}.  Refs.\ \onlinecite{Ha06,Uji} estimated $E_g$ as
$40\div50$ K, so the field of $45$~T may be sufficient to exceed the
threshold (\ref{Delta}) at the ambient pressure.  The experiment can
be also performed in pulsed fields or under pressure, where the anion
superstructure is progressively suppressed \cite{Shinagawa}.
Measurements of the interlayer conductivity using pulsed magnetic
fields of 46 T were performed in $\rm(TMTSF)_2ClO_4$ \cite{Yoshino06},
but the field was applied close to the $x$ axis, rather than to the $y$
axis, as required for our effect.

\begin{figure}
\includegraphics[width=1.0\linewidth]{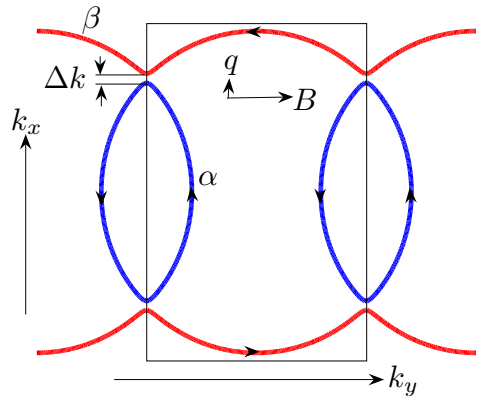}
\caption{(Color online) The in-plane Fermi surface of
  $\kappa$-$\rm(ET)_2Cu(NCS)_2$.  The $\alpha$ and $\beta$ branches of
  the Fermi surface are separated by the distance $\Delta k$ in the
  momentum space.}
\label{fig:pqr}
\end{figure}

A similar analysis can be also applied to the material
$\kappa$-$\rm(ET)_2Cu(NCS)_2$, whose in-plane Fermi surface is shown
in Fig.~\ref{fig:pqr}.  The separation $\Delta k$ between the $\alpha$
and $\beta$ branches of the Fermi surface can be measured by applying
an in-plane magnetic field in the horizontal direction in
Fig.~\ref{fig:pqr}.  This field shifts the Fermi surface of one layer
by the vector $\bm q$ shown in Fig.\ \ref{fig:pqr}.  The threshold
magnetic field, at which the $\alpha$ branch in one layer starts to
touch the $\beta$ branch in the other layer, can be calculated from
Eq.~(\ref{q}).  Using $\Delta k=0.17$~nm$^{-1}$ and the interlayer
distance $c=1.62$~nm \cite{book-Ishiguro,Goddard}, we estimate that
the threshold magnetic field is of the order of 430 T, which is beyond
the current experimental capabilities.

\section{Conclusions}

We have shown that the modifications of the electron dispersion due to
the anion ordering in $\rm(TMTSF)_2ReO_4$ and $\rm(TMTSF)_2ClO_4$
generate effective tunneling amplitudes between many distant chains.
These amplitudes cause peaks in the interlayer conductivity
$\sigma_{zz}$ at many Lebed magic angles (\ref{nm}).  The different
wave vectors of the anion ordering, $\bm Q=(0,1/2,1/2)$ in
$\rm(TMTSF)_2ReO_4$ and $\bm Q=(0,1/2,0)$ in $\rm(TMTSF)_2ClO_4$,
result in the odd and even Lebed magic angles, as observed
experimentally \cite{Kang03,Shinagawa}.  Our theory also explains why
the Lebed oscillations are strong and the DKC oscillations are weak in
$\rm(TMTSF)_2ReO_4$, and vice versa in $\rm(TMTSF)_2ClO_4$, as
observed experimentally \cite{Kang06}.

When a strong magnetic field is applied parallel to the layers and
$B_y$ exceeds a certain threshold, then interlayer tunneling between
different branches of the Fermi surface, produced by folding of the
Brillouin zone, should become possible.  This effect would be observed
as a sharp increase in interlayer conductivity.  It can be utilized
for a direct measurement of the anion gap $E_g$.  Theoretical
description of this effect required a quantum-mechanical treatment of
the wave functions confined to different layers and cannot be achieved
within the framework of quasiclassical electron orbits on a warped
Fermi surface.

Experimental observation of the high number of magic angles (up to 21
in Ref.~\onlinecite{Kang03}) demonstrates a very high level of quantum
coherence achieved in the Q1D organic conductors at low temperatures.
This is remarkable given that the $\rm(TMTSF)_2X$ materials have
strong electron interactions.  In different parts of their rich phase
diagram, these materials have the Mott insulating phase and other
exotic phases \cite{book-Ishiguro,book-Lebed}.  It would be very
interesting to study what happens to AMRO when the system is driven
toward the Mott state using pressure or other variables.

We point out that the theory of the angular magnetoresistance
oscillations (AMRO) in Q1D conductors is equivalent to the
mathematically description of the Mach-Zehnder interference in a
driven superconducting qubit
\cite{Oliver05,Berns06,Sillanpaa06,Izmalkov08,Ashhab07} and of laser
cooling in ion traps \cite{DeVoe89}.  The similarity in the behavior
of these systems demonstrates that quantum coherence in the Q1D
organic conductors at low temperatures is as high as in the
superconducting qubits and ion traps, which are actively considered
for applications in quantum computing and quantum information.  Thus,
the physics of Q1D conductors may have applications in quantum
engineering well beyond the domain of solid-state material science.

\begin{acknowledgments}
V.M.Y. is grateful for the discussions with W.~Kang, S.~Hill, M.J.~Naughton, S.~Uji, W.D~Oliver, and S.~Ashhab, and for the e-mail communications with A.G.~Lebed and L.~Levitov.
\end{acknowledgments}

\appendix

\section{Calculation of the matrix elements}
\label{App:MatrixElements}

In this appendix, we calculate the matrix elements of interlayer
tunneling introduced in Eq.~(\ref{sigma_zz}).

In the case of $\rm(TMTSF)_2ClO_4$, the interlayer tunneling with the
amplitude $t_c$ occurs between the chains of the same type, as shown
in Fig.~\ref{fig:lattice}(b) \cite{t_c'}.  The in-plane Hamiltonians
of two adjacent layers are given by the same expression,
\begin{equation}
  \hat H = \left( \begin{array}{cc}
  E_g  & 2t_b\cos(k_y b) \\
  2t_b\cos(k_y b) & -E_g
  \end{array} \right).
\label{in-plane}
\end{equation}
The eigenvalues $\lambda_\pm$ and the eigenvectors $|\psi_\pm\rangle$
of the Hamiltonian (\ref{in-plane}) are
\begin{eqnarray}
  && \lambda_\pm = \pm \sqrt{[2t_b\cos(k_yb)]^2 + E_g^2},
\label{lambda}\\
  && |\psi_\pm(k_y)\rangle = \frac{1}{N_\pm}\,
  (\lambda_\pm+E_g, 2t_b\cos k_yb),
\\
  && N_\alpha = \sqrt{[2t_b\cos(k_yb)]^2 + (\lambda_\alpha+E_g)^2}.
\end{eqnarray}
The matrix elements of tunneling are proportional to the scalar
products of the wave functions in adjacent layers:
\begin{eqnarray}
  M_{--} &=& \langle\psi_-(k_y+q_y)|\psi_-(k_y)\rangle,
\\
  M_{++} &=& \langle\psi_+(k_y+q_y)|\psi_+(k_y)\rangle
\end{eqnarray}
for tunneling between the same kinds of bands and
\begin{equation}
  M_{-+} = \langle\psi_-(k_y+q_y)|\psi_+(k_y)\rangle
\label{+-}
\end{equation}
between different kinds of bands.  It is clear from Eq.~(\ref{+-})
that $M_{-+}$ vanishes for $q_y=0$ because $|\psi_+(k_y)\rangle$ and
$|\psi_-(k_y)\rangle$ are orthogonal.

In the case of $\rm(TMTSF)_2ReO_4$, the inter-layer tunneling with the
amplitude $t_c$ occurs between the chains of different types.  The
in-plane Hamiltonian of one layer has the form (\ref{in-plane}),
whereas the sign of $E_g$ is reversed in the Hamiltonian $H'$ of
another layer
\begin{equation}
  H' = \left( \begin{array}{cc}
  -E_g  & 2t_b\cos(k_y b) \\
  2t_b\cos(k_y b) & E_g
  \end{array} \right).
\end{equation}
The eigenvalues of $H'$ are the same as in Eq.~(\ref{lambda}), but the
corresponding eigenvectors are different,
\begin{eqnarray}
  && |\psi_\pm'(k_y)\rangle = \frac{1}{N_\pm'}\,
  (\lambda_\pm-E_g, 2t_b\cos k_yb),
\\
  && N_+' = N_-, \qquad N_-' = N_+.
\end{eqnarray}
The scalar products of the wave functions in the adjacent layers now are
\begin{eqnarray}
  M_{--} &=& \langle\psi_-'(k_y+q_y)|\psi_-(k_y)\rangle,
\\
  M_{++} &=& \langle\psi_+'(k_y+q_y)|\psi_+(k_y)\rangle
\end{eqnarray}
for the same kinds of bands and
\begin{equation}
  M_{-+} = \langle\psi_-'(k_y+q_y)|\psi_+(k_y)\rangle
\label{+-'}
\end{equation}
 for different kinds of bands.
Now $M_{-+}$ does not vanish for $q_y=0$, because
$|\psi_+(k_y)\rangle$ and $|\psi_-'(k_y)\rangle$ are not orthogonal.



\begin{thebibliography}{00}
\bibitem{error} Eq.~(12) in Ref.~\onlinecite{Cooper06} gives a similar
  expression for $\sigma_{zz}$ in the case of multiple tunneling
  amplitudes $t_l$.  However, it contains an error and should be
  replaced by the correct Eqs.~(\ref{sigma}) and (\ref{Anl}) given
  here.
\bibitem{t_c'} In Secs.~\ref{Sec:ReO4} and \ref{Sec:Interband}, we
  ignore the tunneling amplitude $t_c'$ introduced in
  Sec.~\ref{Sec:PF6} and consider only the main amplitude $t_c$.  The
  effects discussed in Secs.~\ref{Sec:ReO4} and \ref{Sec:Interband}
  can be obtained already at $t_c'=0$.
\end{thebibliography}
\end{document}